\documentclass[a4paper]{llncs}

\usepackage{color,calc,graphicx,subfig}
\usepackage{amsmath}
\usepackage{amsfonts}
\newcommand{\one}{\mathrm{I} \! \! 1}
\usepackage{amssymb}
\usepackage{amstext}

\newtheorem{prop}{Observation}\def\PRO{\begin{prop}}\def\ORP{\end{prop}}
\newtheorem{theo}{Theorem}\def\TH{\begin{theo}}\def\HT{\end{theo}}
\newtheorem{defi}{Definition}\def\DE{\begin{defi}}\def\ED{\end{defi}}

\usepackage{amssymb}
\usepackage{graphicx}

\usepackage{url}
\urldef{\mailsa}\path|{alfred.hofmann, ursula.barth, ingrid.haas, frank.holzwarth,|
\urldef{\mailsb}\path|anna.kramer, leonie.kunz, christine.reiss, nicole.sator,|
\urldef{\mailsc}\path|erika.siebert-cole, peter.strasser, lncs}@springer.com|

\begin{document}

\mainmatter

\title{Entanglement, Flow and Classical Simulatability \\ in \\ Measurement Based Quantum Computation}

\author{Damian Markham\inst{1} \and Elham Kashefi\inst{2}}

\institute{CNRS LTCI, D\'{e}partement Informatique et R\'{e}seaux, Telecom ParisTech, 23 avenue d'Italie, CS 51327,  75214 Paris CEDEX 13, France \and
School of Informatics, University of Edinburgh, 10 Crichton Street, Edinburgh EH8 9AB, UK}

\toctitle{Entanglement and Flow}
\tocauthor{Damian Markham (CNRS, Telecom ParisTech), Elham Kashefi(University of Edinburgh)}
\maketitle

\begin{abstract}
The question of which and how a particular class of entangled resource states (known as graph states) can be used for measurement based quantum computation (MBQC) recently gave rise to the notion of \emph{Flow} and its generalisation \emph{gFlow}. That is a causal structure for measurements guaranteeing deterministic computation. Furthermore, gFlow has proven itself to be a powerful tool in studying the difference between the measurement-based and circuit models for quantum computing, as well as analysing cryptographic protocols. On the other hand, entanglement is known to play a crucial role in MBQC. In this paper we first show how gFlow can be used to directly give a bound on the classical simulation of an MBQC. Our method offers an interpretation of the gFlow as showing how information flows through a computation, giving rise to an information light cone. We then establish a link between entanglement and the existence of gFlow for a graph state. We show that the gFlow can be used to bound the entanglement width and what we call the \emph{structural entanglement} of a graph state. In turn this gives another method relating the gFlow to bounds on how efficiently a computation can be simulated classically. These two methods of getting bounds on the difficulty of classical simulation are different and complementary and several known results follow. In particular known relations between the MBQC and the circuit model allow these results to be translated across models.
\end{abstract}

Measurement Based Quantum Computing (MBQC) \cite{RB01} has attracted attention recently for its potential towards the realisation of a quantum computer, its role in understanding the power and significance of entanglement for computation \cite{AB09,HWB11}, and that it plays a key role in the development of cryptographic protocols \cite{BFK09,MS08}.
In MBQC one starts off with a large multiparty entangled resource state and the computation is driven by a series of local measurements, the choice of which can depend on the result of previous measurements in the series. The formal language for MBQC was jointly developed by Prakash Panangaden in \cite{Mcal07}. In this work we are interested in the question of how to recognise or characterise a `good' resource for measurement based quantum computing. Given the fact that after the generation of the state, all operations are local, it is natural to expect entanglement to play a key role. Indeed it has been shown that the entanglement of a resource state must be sufficiently high for it to be universal and not classically simulatable (we note that these two properties are currently not known to be equivalent, though it is broadly expected that they are) \cite{Vidal03b,Yoran06,VandenNest06,VandenNest06b,VandenNest07}.

A related question to universality is that of the ability, or not, for a resource to allow any unitary MBQC computation on it at all. This question is addressed by what is called Flow or its generalisation referred to hereinafter as gFlow \cite{Danos06,Browne09}, for a particular class of resource states, called graph states \cite{HEB04} and its extension open graph state \cite{Danos06} (see also below). There exist efficient algorithms \cite{Beaudrap06,MhallaP08} to find gFlow if it exists, and once gFlow is found, it gives an explicit measurement pattern which gives a unitary computation across the resource graph state in hand. Subsequently gFlow has been a useful tool for exploring many aspects of MBQC such as efficient translation between MBQC and the circuit model \cite{Danos06,Beaudrap09}, analysing cryptographic protocols \cite{KMMP09}, direct pattern design in MBQC \cite{BDKR08}, proving bounds on depth complexity \cite{Browne09,BK09} and from a more fundamental perspective, the arrival of causal order in MBQC \cite{BK09,daSilva11,Raussendorf11,Raussendorf12}.

In this paper we show that gFlow also gives a bound on the difficulty in classically simulating MBQC, and how it can be interpreted as a flow of information. This leads to the observation that the \emph{causal forward cone} (the `forward cone' of a qubit is given by the qubits who's corrections directly or indirectly depend on that qubit's measurement results) is equal to the information cone (the cone of qubits where the information spreads to through the computation). We then establish an intuitive link between gFlow on the one hand, and entanglement of a resource state on the other. We further make this connection explicit by showing how gFlow can be used to give bounds on the entanglement of a graph state. In this way we will see that properties of simulateability of MBQC on a resource in terms of entanglement can be translated to conditions in terms of gFlow. Via a known relationship between the circuit model and MBQC these results can also lead to conditions on simulateability of circuits. One such example is a rederivation of the result by Jozsa \cite{Jozsa06}.

The organisation of this papers is as follows. In Section \ref{SEC: End, Det and MBQC} we mention basic observations about entanglement conditions for any good resource state for MBQC which will be then linked to gFlow. In Section \ref{SEC: gFlow} we introduce graph states and review the notion gFlow and several preliminary notions necessary for the rest of the paper. In Section \ref{SEC: Direct simulation} we prove that gFlow can be used to give bounds on direct simulation of a MBQC. In Section \ref{SEC: InfoFlow} we discuss how gFlow can be used to see how information flows through a resource in an MBQC, giving in particular an information light cone which coincides with the causal cone as defined in  \cite{Raussendorf11,Raussendorf12}. In Section \ref{SEC: Ent Simulation Bound} we show how gFlow can be used to bound the entanglement of a resource state which gives a new route to bounding simulatability of a MBQC, which is different and complementary to the direct simulation in Section \ref{SEC: Direct simulation}. We finish with discussions.

\section{Entanglement and Determinism in MBQC}\label{SEC: End, Det and MBQC}

In measurement-based quantum computing one starts with a large entangled resource state $|\Psi_{RES}\rangle$ on $n$ qubits.
We identify a two sets of qubits, $I$ which will represent the inputs, and $O$ which will represent the outputs of the computation, with $n\geq|O|\geq|I|$. Generally one can consider three types of computation using this resource, one with a classical input and a classical output (let's call this $CC$), one with a quantum input and a classical output $QC$ and one with a quantum input and a quantum output $QQ$. Clearly $QQ$ is the most general, since one can always encode classical information onto quantum states. In this work we focus on $QQ$. When considering a quantum input $|\psi\rangle_S$ (on a system $S$ of $|I|$ qubits) the first step is to teleport the input system qubits $S$ onto $I$ on the resource state, by some global map on $I$ and $S$. This can be done for example by entangling $I$ with $S$ (using, say, a control-$Z$ gate) and performing Pauli $X$ measurements on $S$ then appropriate corrections (see e.g. \cite{MMP13} for graph state resources).
The computation then proceeds by a series of measurements on individual qubits, followed by corrections, then further measurements and corrections and so on until the computation is complete. We call the sequence of measurements and corrections the \emph{measurement pattern} (see \cite{Mcal07} for formal definitions). The outputs qubits, labelled $O$ are those qubits which at the end of the computation are not measured. In this way the computation uses the resource state to transfer the input from $I$ to $O$, in a kind of involved teleportation, at the same time performing some unitary over the input.

We begin with the following definitions.

\DE \label{d-unit} A resource state $|\Psi_{RES}\rangle$ on $n$ qubits, with defined input qubits $I$ and output qubits $O$ is \emph{D-Happy} if for all bi-partitions $A,B$ such that $I\in A$ and $O \in B$ we have
\begin{eqnarray}
E_{A,B}(|\Psi_{RES}\rangle)\geq |I|,
\end{eqnarray}
where $E_{A,B}(|\Psi_{RES}\rangle)=S(\rho_A)=S(\rho_B)$ is the entropy of entanglement across partition $A,B$, where $S(\rho_A)=-\ln Tr(\rho_A)$ is the von Neumann entropy of the reduced state $\rho_A = Tr_B(\rho_{AB})$.
\ED

\DE
A MBQC pattern is called \emph{unitary} if for all inputs the returned state of the output is an encoding of a unitary acting on the input.
\ED

A similar notion was defined in \cite{Mhalla11} as \emph{information preserving pattern}. We present a simple but important observation about the link between the above two definitions.

\PRO \label{t-dhappy} There exists a unitary MBQC pattern on a resource state $|\Psi_{Res}\rangle$ on $n$ qubits, with input qubits $I$ and output qubits $O$ only if it is D-Happy.\ORP

\noindent {\bf Proof.} We start by noting that in order to teleport a state $|\psi\rangle \in \mathbb{C}_2^{\otimes |I|}$ perfectly across $|\Psi_{Res}\rangle$ to its output space, it is necessary to have $E_{A,B} \geq |I|$, for all bi-partitions $A,B$ such that $I\in A$ and $O \in B$. To see this is true one can consider the state to be teleported as half a maximally entangled state. After the teleportation one would end up with a state with entanglement $E_{A,B} = |I|$. Since all operations are local, and it is not possible to increase entanglement in the process, this implies that we started with $E_{A,B} \geq |I|$. We then note that any measurement based computation can be considered as a teleportation across any cut which divides the inputs from the outputs - since all operations are local to each qubit, they are certainly local to any cut. $\square$

Recall that a MBQC computation evolves through various branches, depending on the measurement outcomes. In a unitary MBQC pattern as defined above, it is possible that different branch implements different unitary operators. A weaker notion of unitary computation is given below.

\DE \label{d-det} A measurement based quantum computation is called \emph{deterministic} if for all inputs the returned state of the output is an encoding of a fixed unitary acting on the input independent of the branch of the computation.
\ED
Other types of determinism and their connections can be found in \cite{Browne09,Mhalla11}. In this paper we only consider the above central notions as they can be directly linked to the concept of structural entanglement as we present later. Moreover it is known that for graph states with $|I|=|O|$ the two definitions of unitary and determinism, defined above, are equivalent \cite{Mhalla11}. This will allow us to link the concept of gFlow to D-Happiness as we discuss later.

\section{Preliminaries: Graph states, Flow and gFlow} \label{SEC: gFlow}

In the previous section we presented a necessary condition for computation across a resource state based on entanglement. The simple idea there was that if information can be transferred across a resource state, that state must be maximally entangled across each cut. We did not say anything about how this can be done however. This is where the ideas of Flow \cite{Danos06} and its generalisation gFlow\cite{Browne09} play a role, where a constructive definition together with efficient algorithm could be obtained for particular class of resource states of many qubits - graph states (defined below), with chosen input $I$ and output $O$. If a graph state has gFlow, it implies that a unitary computation can be carried out across it \cite{Danos06,Browne09}. Not only that, gFlow also gives instructions how to do it, and tells you what class of computations will be carried out (which unitaries). We will show that we can further use gFlow to give a simple bound on classical simulation of the computation based on the size of the forward cones implied by the measurement patters. This gives rise to an interpretation of the gFlow as showing us how information is `spread' across the resource state throughout the computation, in an information light cone (which coincides with the causal forward cone in MBQC \cite{BK09,Raussendorf11,Raussendorf12}). In this section we review the definitions of graph states \cite{HEB04}, open graph states, Flow \cite{Danos06} and gFlow \cite{Browne09} and related concepts.

We start by defining the resource states considered, graph states \cite{HEB04}. A graph state is a multipartite state $|G\rangle$ of $n$ qubits, in one to one correspondence with a simple undirected graph $G$, with vertices $V$ and edges $E$. Every vertex is associated to a qubit, and every edge can be understood as an entangling operation between qubits which have been initialised in the state
\begin{equation}
|+\rangle:=(|0\rangle+|1\rangle)/\sqrt{2}. \nonumber
\end{equation}
We then have
\begin{equation}\label{EQN: Graph state}
|G\rangle_V := \prod_{i,j\in E} CZ_{i,j} |++ \cdots +\rangle_V,
\end{equation}
where $CZ_{i,j}$ is the control-$Z$ operation between qubits $i$ and $j$. It is clear from this definition that the entanglement across a cut $A,B$ is bounded by the number of edges cutting it, denoted $C_{A,B}$,
i.e. $E_{A,B}\leq C_{A,B}$.

Graph states can equivalently be defined by their stabiliser operators \cite{HEB04}, a set of $n$ operators, each associated to one vertex defined as
\begin{equation}
\label{Eqn: Definition of Stabilisers}
K_i:= X_i \otimes_{j \in N(i)} Z_j,
\end{equation}
where $X$ and $Z$ are the Pauli operators (and $Y=iZX$). The graph state $|G\rangle$ is the unique state satisfying all the eigenvalue equations (also called stabiliser relations or equations)
\begin{equation}
K_i |G\rangle_V= |G\rangle_V. \nonumber
\end{equation}
The above relation is the key to how gFlow works - gFlow tells us how to apply the stabilisers to correct for measurements.

When used as a resource state for MBQC we assign some vertices as inputs $I\in V$ and some as outputs $O\in V$. In order to preserve the space we have that the size of the input set $|I| \leq |O|$. We call the graph, with these assignments an \emph{open graph} denoted as $G(I,O,V)$. The associated state is slightly different, the input vertices are no longer prepared in the $|+\rangle$ state, but can be arbitrary input qubits $|\psi\rangle_I$. The rest of the vertices are prepared as normal, and again, every edge corresponds to a control-$Z$ operation. We denote such a state as $|G(\psi)\rangle$
\begin{equation} \label{EQN: Open graph state}
|G(\psi)\rangle_V := \prod_{i,j\in E} CZ_{i,j} |\psi\rangle_I |+ \cdots +\rangle_{V/I}
\end{equation}
where the state only depends on the inputs so different open graphs may have the same open graph state if they share the same set $I$, and graph $G$ even if they have different assigned outputs $O$. The stabilisers are now reduced to those only on the non-inputs (we denote this set $I^c$)
\begin{align} \label{EQN: stab}
K_i |G(\psi)\rangle_V = |G(\psi)\rangle \;\;\;\;\; \forall i \in I^c.
\end{align}
Here the stabilisers define a space (of dimension $2^{|I|}$) of states such that this equation holds. The open graph state defined in Equation \ref{EQN: Open graph state} is equivalent to starting in the standard graph state Equation \ref{EQN: Graph state} and teleporting an input $| \psi\rangle_S$ over system $S$ (of $|I|$ qubits) onto the input vertices $I$ by performing control-$Z$s between $S$ and $I$, followed by Pauli $X$ measurements on the $S$ qubits and corrections (see e.g. \cite{MMP13}).

In the standard model of MBQC \cite{Mcal07,Browne06} measurements are performed in one of the equatorial planes defined by the $X-Y$, $X-Z$ or $X-Y$ planes, and correction operations are local Pauli operators. By the end of the computation all vertices will be measured except the outputs (we denote this $O^C$). The gFlow assigns a set of correction operators for each of these measurements.

Before giving the definition of gFlow, we give the intuition to how it works for measurements in the $X-Y$ plane. This corresponds to measuring in the basis $|\pm^\theta\rangle:= (|0\rangle \pm e^{i\theta} |1\rangle)/\sqrt{2}$. We denote the projections associated to results $\pm1$ as $P^{\pm,\theta}:=|\pm^\theta\rangle\langle \pm^\theta|$. For later use we denote the results in binary form as $r_i=0$ for $+1$ and $r_i=1$ for $-1$ outcomes. When measureing a state $|\psi\rangle$, in quantum mechanics the result is random (in fact normally in MBQC the probabilities are $1/2$ and $1/2$), which takes the resulting state to one of two branches, either the positive branch $P^{+,\theta}|\psi\rangle/p_+$ with probability $p_+= \langle \psi | P^{+,\theta}|\psi \rangle$, or the negative branch $P^{-,\theta}|\psi\rangle/p_-$ with probability $p_-= \langle \psi | P^{-,\theta}|\psi \rangle$.

Clearly to perform a deterministic computation $U$, we need to recover a deterministic evolution, hence corrections need to be applied. By convention we take the positive branch to be the ideal branch (note that of course $P^{+,\theta}|\psi\rangle/p_+$ is not in general a unitary embedding, this is an additional requirement which is also satisfied for our case). The task is then to find a correction operator to take the state when projected onto the $-1$ result to that of the $+1$ result (possibly ignoring the state of the measured qubit, since it is no longer used). The starting point is to notice that for all measurements in the $X-Y$ plane, the projections are related to each other by a Pauli $Z$ operator (for the other planes it is similarly the orthogonal Pauli operator) $P^{+,\theta} = Z P^{-,\theta} Z$. Imagine if it were possible to know the outcome of the measurement before it was performed (for example by traveling back in time after the measurement was performed and telling yourself), instead of correcting after the event, if we knew that we were about to get $-1$, we could cheat and apply a Pauli $Z$ operator - then the `measurement' (projection) would take us onto the projection we wanted, the positive branch. Obviously this is not possible without time travel since in quantum mechanics the results of measurements are random and cannot be known beforehand (we can only predict probabilities). However, we can use the stabilisers to \emph{simulate} this strategy.

Imagine we applied the measurement on qubit $i$, then our time-travelling correction strategy for the $-1$ result would be to perform a Pauli $Z$ operator on qubit $i$. Now, if we take a neighbour $j\notin I$, the stabiliser condition (Equation \ref{EQN: stab}) tells us that
\begin{align}
Z_i |G(\psi)\rangle &= Z_i K_{j=N(i)} |G(\psi)\rangle  \label{EQN: stab correction}\\
&= \one_i \otimes X_{j}\otimes_{k\in N(j)\neq i}Z_k |G(\psi)\rangle.\label{EQN: stab correctionII}
\end{align}
Since $X_{j}\otimes_{k\in N(j)\neq i}Z_k$ are on different systems from the measured qubit $i$, it does not matter when they are performed (they commute with the measurement). In this way, applying $X_{j}\otimes_{k\in N(j)\neq i}Z_k$ correction operator after the measurement, is the same as applying a $Z$ correction before the measurement - so that it has exactly the same effect. The later is sometimes called an `anachronical correction', since it is as if we could go back in time and correct the measurement before it happened. The same works if a product of stabilisers is used in Equation \ref{EQN: stab correction} as long as their product results in one Pauli $Z$ operator on qubit $i$, and we call the vertices associated to these stabilisers as the \emph{correcting set}. Graphically this condition is ensured if the total number of edges between the correcting set and the vertex being measured is $1$ modulo $2$. This motivates the definition of the odd neighbourhood of a set of vertices $K$, denoted $Odd(K):=\{\mu \;| \;\; |N(\mu) \cap K|=1 \;\text{mod}\;2 \}$, which will be used in the definition of gFlow below.

Using this idea, gFlow plays the role of making sure it is possible to make a good choice of which neighbour (or set of neighbours) to choose in a consistent way - so that corrections do not somehow contradict or interfere with one another. Indeed, gFlow is composed of a time order $\prec$ (partial order over vertices) and a choice of neighbouring sets (correcting sets) for each measured vertex $i$, denoted $g(i)$ with this in mind. Firstly the time order should be consistent, so that corrections happen after the assigned measurements - this appears as \emph{(g1)} in the Definition \ref{DEF: gFlow} below. Secondly, the correction should not invalidate or effect earlier corrections. This is true if no Pauli $Z$ operators appear in the past when applying the stabiliser corrections, i.e. the correcting set is not oddly connected to the past - this appears as \emph{(g2)} in Definition \ref{DEF: gFlow}. Finally the correcting set should correct for the measurement it is assigned to. For measurements in the $X-Y$ plane this corresponds to the application of a Pauli $Z$ operator when the correcting stabilisers are applied, which means the correcting set should be oddly connected to the measured vertex - which appears as \emph{(g3)} in the definition below (the analogous corrections for the other planes appear after).

\DE \label{g-flow} \label{DEF: gFlow} An open graph state $G(I,O,V)$ has gFlow if there exists a map $g:O^c \rightarrow \mathcal{P}^{I^c}$ (from measured qubits to a subset of prepared qubits) and a partial order $\prec$ over $V$ such that for all $i\in O^c$ \\
(g1) if $j \in g(i)$ and $i\not=j$ then $i\prec j$\\
(g2) if $j \prec i$ and $i\neq j$ then $j \notin Odd(g(i))$ \\
(g3) for measurements in the $X-Y$ plane, $i \notin g(i)$ and $i \in Odd(g(i))$ \\
(g4) for measurements in the $X-Z$ plane, $i \in g(i)$ and $i \in Odd(g(i))$ \\
(g5) for measurements in the $Y-Z$ plane, $i \in g(i)$ and $i \notin Odd(g(i))$ \\
Flow is a special case of gFlow, when all measurements are performed on the $X-Y$ plane, and the correction sets $g(i)$ have only one element.
\ED

In this way the product of $\prod_{j\in g(i)} K_j $ applies the appropriate `anachronical' correction on vertex $i$, whilst not effecting other previous corrections. The associated computation can be carried out as follows. First generate the open graph state, then go through round by round (in the order given by $\prec$), measureing each qubit $i$, denoting the binary form of the outcome $r_i$, followed by the correction given by
\begin{equation} \label{EQN: gFlow Correction}
\left(\sigma_i \prod_{j\in g(i)} K_j \right)^{r_i}
\end{equation}
where $\sigma_i$ is the pauli $Z_i$, $Y_i$ or $X_i$ for measurement on qubit $i$ done on the $X-Y$, $X-Z$ or $Y-Z$ planes respectively, so that Equation \ref{EQN: gFlow Correction} is trivial over $i$ and non-trivial only on future qubits of $i$ i.e. on $j$ such that $i\prec j$.

\begin{figure}[h]
\centering
{\resizebox{!}{4.3cm}{\includegraphics{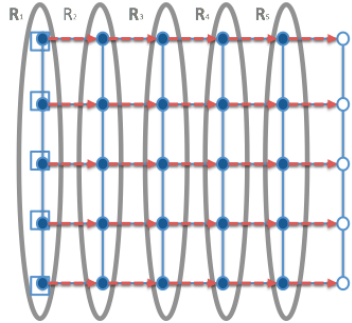}}} \caption{\label{FIG: gFlowCLUSTER} An example of gFlow for the two dimensional clusters state as an open graph state. Following the convention in \cite{Danos06,Browne09} inputs are identified by vertices with squares around them, and outputs are identified as hollow vertices (hence all non-hollow vertices will be measured in the computation). The choice of gFlow for a given vertex is indicated by red dotted arrows from the vertex to its gFlow (these are called \emph{gFlow paths}, see Definition \ref{gPath}. Note that gFlow paths need not follow graph edges, as in Figure~\ref{FIG: gFlowEGs}b). The induced measurement rounds are highlighted in grey, (see Definition \ref{Round}).}
\end{figure}

In \cite{Browne09} it is shown that gFlow is a necessary and sufficient condition for an open graph state to allow a \emph{uniform} unitary, deterministic computation to be performed across it, where uniform means that each qubit can be measured at an arbitrary angle on one of the planes.  Hence the existence of gFlow implies the resource is also D-Happy. Intuitively on can think that the existence of gFlow guarantees that the entanglement of the graph state is such that the random effects of local measurements can be absorbed and countered by yet unmeasured qubits. The following definitions will be used to discuss how information travels throughout the computation \cite{BK09}.

\DE \label{gPath} A \emph{gFlow path} starting from a vertex $\mu$, denoted as $gPath(\mu)$, is an ordered set of vertices such that for each pair $(i,j)$ we have $j\in g(i)$ and the first element of the set is $\mu$.
\ED
\DE \label{gInf} An \emph{influencing path} starting from a vertex $\mu$, denoted as $gInf(\mu)$, is an ordered set of vertices such that each pair $(i,j)$ is on a gFlow path or is preceded immediately by a pair on a gFlow path.
\ED
\DE \label{Forward Cone} The \emph{forward cone} $F_C(\mu)$ of a vertex $\mu$ is the set of all vertices touched by all influencing paths from $\mu$.
\ED
The concept of the forward cone appears in \cite{BK09,Raussendorf11,Raussendorf12} and can be understood as a causal light cone, as described in \cite{Raussendorf11,Raussendorf12}. The partial order $\prec$ in a gFlow defines time order for the rounds of measurements. We say a vertex $\mu$ is in a round $R_x$ if it is measured in round $x$.
\DE \label{Round} The set $R_x$, denotes the set of vertices which are measured in the $x$th round of measurements according to the $gFlow$.
\ED

\begin{figure}[h]
\centering
{\resizebox{!}{4.3cm}{\includegraphics{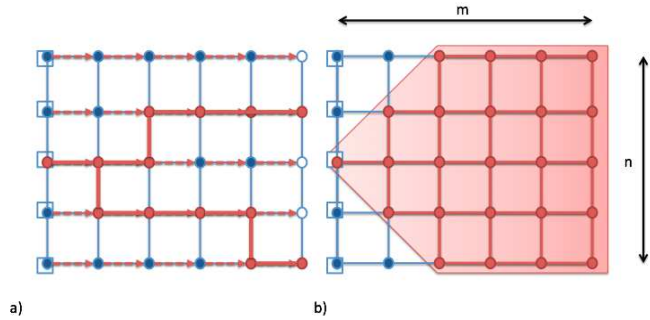}}} \caption{\label{FIG: ClustergFlowVol} a) The bold red lines are examples of two possible influencing paths from the central input vertex (see Definition \ref{gInf}), for the gFlow paths given by the red dotted arrows. An influencing path is path which follows gFlow paths and no more than one edge between gFlow paths. b) The collection of all influencing paths identifies the set of vertices (in red) in the \emph{forward cone} (see Definition \ref{Forward Cone}). The maximum size of forward cone for the 2D cluster state is indicated by the red shaded region (for the same gFlow). For an $n \times m$ 2D cluster state the maximum forward cone is of size $|F_{C_{max}}|=nm-n^2/4$. This gives a bound on classical simulation for a computation, in Observation \ref{OBS: gFlow simulation}. The same region has an interpretation as an information light cone (see Section \ref{SEC: InfoFlow}). }
\end{figure}

The best way to understand these definitions is through some examples. The gFlow (which is also a Flow in this case) is illustrated for the 2D cluster state in Figure \ref{FIG: gFlowCLUSTER}. In Figure \ref{FIG: ClustergFlowVol} we show examples of influencing paths and their union, which  make up the forward cone for the 2D cluster state.

\begin{figure}[h]
\centering
{\resizebox{!}{4.3cm}{\includegraphics{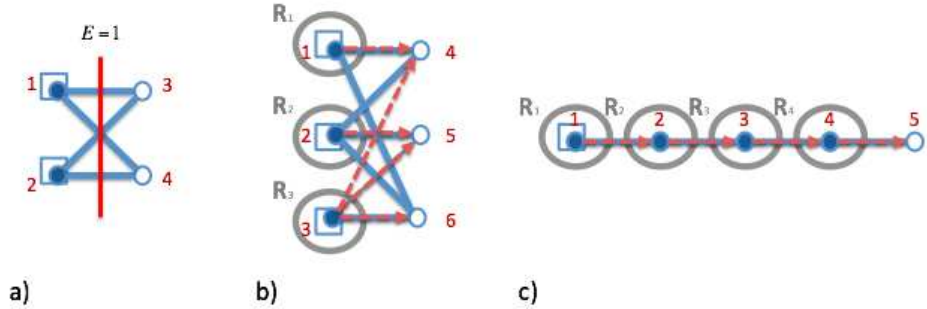}}} \caption{\label{FIG: gFlowEGs} Examples of open graph states with and without gFlow. The  gFlow paths are red dotted lines, and the induced measurement rounds are highlighted in grey (see Definition \ref{Round}). The graph in Figure a) does not have gFlow. This can be seen since the entanglement across the cut input/output is lower than the number of inputs. The graph in Figure b) has gFlow but no Flow \cite{Browne09}. The graph in Figure c) is the linear cluster state which has a gFlow that is also Flow.}
\end{figure}

Before moving on to the interpretations of gFlow with respect to simulation and information flow, we review some examples which illustrate its power as a tool for analysing entanglement (as potential resources for MBQC), and in accessing the tradeoff between classical processing and number of measurement rounds (depth \cite{BK09}). We start with an example of an open graph for which there is no gFlow in Figure \ref{FIG: gFlowEGs} a). It can easily be seen that there is no possible assignment of correction sets $g(i)$ and time order satisfying the conditions in gFlow for any measurement axes. Indeed its inability to act as a resource for computation across it follows directly from the fact that the entanglement across it is less then the number of inputs (hence it is not D-happy). We note however that there are examples of graph states which are D-happy, but do not allow a gFlow \cite{Browne09,Mhalla11}. All such known examples still do allow computation across them. The second example is one where there exists a gFlow, but it necessarily has some correction sets which have more than one member - i.e. there is no Flow, as shown in Figure \ref{FIG: gFlowEGs} b). The associated gFlow is give by assignments $g(1)=4$, $g(2)=5$ and $g(3)=4,5,6$, with partial order given by the ordered measurement rounds $R_1 = 1$, $R_2=2$ and $R_3=3$, and all measurements in the $X-Y$ plane. Note here that a gFlow path need not lie on a graph edge as for the gFlow path $(3,4)$. The third example is the simple linear cluster state in Figure \ref{FIG: gFlowEGs} c), where gFlow follows along the line.

\begin{figure}[h]
\centering
{\resizebox{!}{4.3cm}{\includegraphics{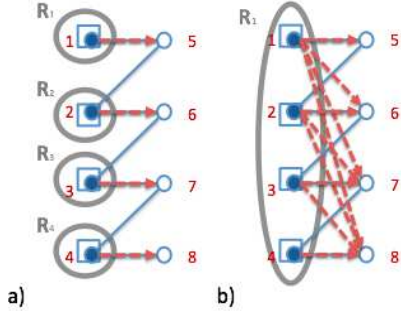}}} \caption{\label{FIG: flowVsgFlowEG} This open graph state has several possible gFlows, and illustrates how gFlow can be used to find advantages in terms of the number of rounds needed (depth) in a computation. a) is a gFlow with one correcting vertex per qubit, hence it is also a flow. This requires a number of rounds scaling with the number of inputs. b) is a gFlow which has largest size scaling with the number of inputs, but all measurements can be done in one round. Indeed all intermediary tradeoffs are also possible.This exemplifies the tradeoff between classical computation required and the number of measurement rounds needed.}
\end{figure}

A final example illustrates how gFlow can be used to find advantages in the number of rounds needed in a computation (taken from \cite{Browne09}). In Figure \ref{FIG: flowVsgFlowEG} the same open graph can have different gFlows. In the first case, Figure \ref{FIG: flowVsgFlowEG} a), the gFlow has correcting sets of size one, hence it is a Flow ($g(i)=i+4$), and the number of rounds is the number of inputs (in the example this is four, but it easily extends to arbitrary size). More complicated gFlows can be found by increasing the size of some correcting sets, with the benefit of reducing the number of rounds. Figure \ref{FIG: flowVsgFlowEG} b), we set the correcting sets as $g(1)=5,6,7,8$, $g(2)=6,7,8$ , $g(3)=7,8$ and $g(4)=8$. It can easily be checked that this assigment allows all measurements to be done in the same round since for every vertex $i$, the correcting set $g(i)$ is oddly connected only to $i$.

\bigskip

The above example illustrates a general scheme that could be understood as a tradeoff between rounds of computation and the amount of classical processing needed, but we have not yet talked about classical processing. To see how it works, we should think back again at what the gFlow does. Recall that gFlow tells us on which sets of vertices we should apply corrections (Equation \ref{EQN: gFlow Correction}). In particular, for a vertex $i$, the correction associated to its measurement result ($r_i$, where $r_i=0$ corresponds to the ideal branch and $r_i=1$ to that which needs to be corrected) is the application of the product of the stabilisers of all the vertices in $g(i)$ (minus the $Z_i$) - i.e. the correction is $(Z_i\prod_{j\in g(i)}K_j)^{r_i}$. Thus, if a vertex $l$ is in the gFlow (or is a neighbour to a gFlow vertex) of another vertex $i$, then it will receive an $X_l^{r_i}$ (or $Z_l^{r_i}$) correction. The total number of corrections for a vertex depends on how many gFlow set (or neighbourhoods of gFlow set) to which that vertex belongs to. In the example Figure \ref{FIG: flowVsgFlowEG} b), vertex $8$ has corrections from all inputs - hence it must receive the correction $X_8^{r_1 \oplus r_2 \oplus r_3 \oplus r_4}$ (where $\oplus$ is the sum mod 2). In general, to calculate the Pauli $X$ correction that should be applied on qubit $j$ requires calculating the parity of all the $r_i$s where $j \in g(i)$ and for the Pauli $Z$ correction the parity of all the $r_i$s where $j$ is a neighbour of $g(i)$. We assume this is done classically (since it is a simple calculation), however, by increasing the size of the gFlows (in order to reduce the depth), we necessarily increase the size of this classical computation. This tradeoff has recently been translated to a tradeoff between the degree of the initial Hamiltonian and time of computation in the adiabatic model \cite{AMA13}.

This tradeoff, a particular feature of the measurement based model, gave rise to a distinction in the power of measurement based quantum computation compared to the circuit model with respect to the number of time steps required \cite{BKP10}.
The first example of a depth separation between quantum circuit and MBQC was proven for the calculation of parity function (where depth is defined to be the minimum number of rounds for a computation) \cite{BK09}. Indeed this is a general feature that the depth of MBQC can be logarithmically better than the circuit model, where the difference is absorbed into the classical processing. More concretely it was shown that the depth complexity of MBQC is equal to the depth complexity for the circuit model with the addition of unbounded fan out gates \cite{BKP10}.

\section{Direct Simulation from gFlow} \label{SEC: Direct simulation}

We will now see how we can derive a simple classical simulation, by tracking the stabilisers and logical operators. This idea is exactly how one can understand the Gottesman Knill theorem for the efficient classical simulatability of computations including only Clifford operations \cite{Gott97}. The proof follows from tracking stabilisers operator since they are an efficient way to describe a stabiliser state (such as a graph state), and Clifford operations, by definition transform stabiliser states to other stabiliser states, so computations can be simply tracked and described \cite{Gott97}.

In what follows we will represent the computation in terms of the evolution of a set of logical operators. In physics there are two main, equivalent, ways that one views quantum evolution. One method (more common in quantum information) is where we look at how a state develops, and keep track of it as it evolves. This is known as the Schr\"{o}dinger representation. Equivalently, one can view the state as having not altered, but the operators defining measurements having changed. This picture is known as the Heisenberg representation of evolution. In between these two pictures lies another way of representing evolution, which has been developed for quantum information - the so called `logical Heisenberg' representation \cite{Gott97,Gott98}. In this method we track the evolution of a complete set of logical operators - in this case the Pauli operators. To recover the Shr\"{o}dinger representation, we remember that any state density matrix can be decomposed into Pauli operators (see Equations (\ref{EQN: density Matrix} and \ref{EQN: density Matrix Update} in the next section). The logical Heisenberg representation has proved a very instructive way to view the evolution of MBQC \cite{RBB03,Browne06}, and as we will see leads to a simple bound on the cost of classical simulation.

Our simulation will follow the main treatment of \cite{RBB03,Browne06}, with the addition that we will consider rotated operators and their decomposition into Pauli operators, and we will use gFlow to instruct our procedure for updating the operators, which eventually leads to our main Observation. Our main tool will be the stabiliser formalism \cite{Gott97}. As mentioned in Section \ref{SEC: gFlow}, for an open graph state the stabilisers define a space. Generally we talk in terms of a stabiliser group $S$, which is a subgroup of the Pauli operators. In the case of the open graph states, the generators of the stabiliser group are given by the operators $K_i$ (Equation \ref{Eqn: Definition of Stabilisers}), so that $S=\langle \{K_i\}_{i=1}^{n}\rangle$. These are not the unique generators, indeed multiplying each of these by any one generator gives a new set of generators. The stabiliser group defines a space (the stabiliser space, or `code' in error correction terminology) by a set of eigen equations - it is the space of states which are unchanged by the group. For the open graph states this is given by Equation \ref{EQN: stab} that is for all $i\in I^c : K_i |G(\psi)\rangle = |G(\psi)\rangle$, which implies all products of $K_i$ (i.e. all elements of S) leave the states unchanged. We say the states $|G(\psi)\rangle$ are \emph{stabilised} by the group $S$. In general if the stabiliser group for $n$ qubits is generated by $k$ elements, then the stabiliser space is of dimension $2^{n-k}$. Essentially the stabilisers act like the identity over the this space, defining the space itself. In addition to tracking the logical operators, we will also track the stabiliser operators - indeed this will be a key tool for the former.

One can picture the whole of the computation in a high level as follows. The attaching of the input to the graph state (forming the open graph state), encodes the input space onto the many qubit state. The information is in some sense `spread' over the large entangled state (we will talk more about computation as spreading of information in the next section). We call this encoded space the logical space. During the computation the information is pushed forward through the measurements towards the outputs, so that after the final measurements the logical space sits on only the output qubits. During this push the logical space is also rotated around, resulting in unitary computation. One can think of the stabilisers as keeping track of where the logical space is sitting, and the logical operators as telling you how the space has been rotated (in a sense the logical operators track both).

If state $|\psi\rangle$ in the stabiliser space, with stabiliser group $S=\{S_i\}$ evolves under unitary $U$, the new state $U|\psi\rangle$ is clearly stabilised by $\{U S_i U^\dagger\}$, giving the updated stabiliser group. Under measurement things are slightly more complicated. In this work we use only single qubit projective measurements, which we write as two outcome measurements of the form $A_i=P_i^+ -P_i^-$ where $P_i^\pm$ are the projectors onto the $\pm 1$ outcomes where $i$ indicates the qubit measured. As usual we denote $r_i$ as the binary representation of the measurement outcome with $r_i=0$ when the outcome is $+1$ and $r_i=1$ when the outcome is $-1$. If it is possible to find a set of generators such that only one anticommutes with the measurement, call it $S_i$, and the rest commute, the update simply replaces $S_i$ with $-1^{r_i} A_i$. It is not hard to see that this group will stabilise the state after measurement \cite{Gott97}. The projection from the measurement will not change the eigenvalue relation of commuting operators, and the projected state is clearly a $+1$ eigen state of the operator $-1^{r_i}A_i$.
The trick is to find a suitable set of generators allowing for such an update (i.e. such that one and only one anticommutes with the measurement) - which is where the gFlow comes in.

So how should we describe the evolution of our logical operators? We want them to describe the information as it evolves. Talking in terms of pure states (which suffices for our discussion)  if $|\psi\rangle \rightarrow |\tilde{\psi}\rangle$, we want that our logical operators evolve $L \rightarrow \tilde{L}$ so that their expectation is preserved, that is we demand $\langle \psi |L|\psi\rangle = \langle \tilde{\psi} |\tilde{L}|\tilde{\psi}\rangle$. In this way, the new operators $\tilde{L}$ genuinely reflect the information of the evolved space (see \cite{Gott98,RBB03,Browne06} for more details). Under a unitary evolution $|\psi\rangle \rightarrow U |\psi\rangle$, we then have $L \rightarrow U L U^{\dagger}$, clearly satisfying our requirement. For measurements, the trick will be to ensure that the logical operators commute with the measurement operators, in which case, they remain unchanged (measuring commuting observables cannot effect their expectation). The way of doing this will be to multiply by stabilisers - which act as identity on the logical space, so can be introduced without effecting the validity of the logical operators.

We will now see how we can track the evolution of the stabilisers and logical operators through the computation. This will be done in three steps. Note that our procedure does not exactly reflect the step by step process of the computation, as we do not consider corrections, rather it reflects the update as if all measurements had the outcome $+1$ - which is indeed the role of the corrections in the first place. In our discussion below we focus on measurements on the $X-Y$ plane, similar arguments simply apply to the other planes.

\textbf{Step 1:} The first step is to prepare the stabilisers in a form that will allow us to simulate the measurements through the computation more easily. Physically it corresponds to the unitary process of applying the control-$Z$ operators generating the open graph states (Equation \ref{EQN: Open graph state}), followed by simplifying the measurement operator by applying first the appropriate local rotation. The stabilisers of the open graph state are already given in Equation \ref{Eqn: Definition of Stabilisers}. For each input $i$ an informationally complete set of operators is given by the Pauli operators $X_i$, $Z_i$ and $Y_i= iZ_i X_i$. If we know $X_i$ and $Z_i$ we can calculate $Y_i$, hence we concentrate only on these two, and denote them as $L_{X_i}$ and $L_{Z_i}$ as we trace them through the computation. The control-$Z$ operators generating the open graph state is unitary, thus after being attached to the graph the logical operators become $L_{X_i}=X_i \otimes_{j \in N(i)}Z_j $ and $L_{Z_i}=Z_i$ (using the relation $L \rightarrow U L U^{\dagger}$ where $U$ is the control-$Z$ operator, see also \cite{Mcal07}).

Now we want to put these in a form ready to simulate measurements. The idea is based on the fact that a measurement in the $X-Y$ plane is equivalent to first rotating around the $Z$ axis, followed by measurement in the $X$ basis (similar relations are true for the other two planes used). We initialise all the stabilisers and logical operators by doing this rotation, and consider Pauli $X$ measurements afterwards. The resulting state is sometimes called a rotated graph state. At the same time we replace the individual stabilisers by products given by the gFlow. We thus start with stabilisers
\begin{equation}\label{EQN: Stabiliser group Step 1}
S=\left< \left\{S_i:=\prod_{j \in g(i)} K_j^{\theta_j} \right\}_{i \in O_C},\{G_i\}_i\in{O} \right>,
\end{equation}
where $K_i^{\theta_i}:=  e^{i \theta_i / 2 Z_i} K_i e^{-i \theta_i / 2 Z_i} = \cos \theta_i X_i \otimes_{j \in N(i)} Z_j + i \sin \theta_i Z_i X_i \otimes_{j \in N(i)} Z_j$ are the rotated graph state stabilisers and $\theta_i$ is the angle of the measurement for qubit $i$. The set $\{G_i\}_i\in{O}$ are there simply to complete the set of generators in the case that $|I|<|O|$, chosen such that $[G_i,X_j]=0$, $\forall j \notin O$. Such a set can always be found as follows, take an arbitrary set of operators which complete a generating set (note that the operators $S_i$ above are by definition all independent and so can form part of a generating set, then there is always some set of operators in S which complete this set of generators). To ensure commutation relation, we go round by round, starting from $R_1$, we go through each vertex $\nu$ in the round, and check if it commutes with $X_\nu$- and if not we multiply it by $S_i$. These operators are still valid generators and they  commute with all the $X_\nu$ measurements. At the same time, by applying the local unitary Phase rotations, the logical operators are initialised to
\begin{align}
L_{X_i} &= e^{i \theta_i/ 2 Z_i} X_i e^{-i \theta_i / 2 Z_i }  \otimes_{j \in N(i)} Z_j, \nonumber \\
&=  \cos \theta_i X_i \otimes_{j \in N(i)} Z_j + i \sin \theta_i Z_i X_i \otimes_{j \in N(i)} Z_j, \nonumber \\
L_{Z_i} &= Z_i. \label{EQN: Log Op Step 1}
\end{align}

\textbf{Step 2:} The second step is to take the logical operators  to a form which is convenient for measuring $X_\nu$ on all the non-outputs - by making sure that they commute with $X_\nu$ $\forall \nu \in O_C$. This update does not actually reflect any physical operation, rather it is just rewriting by multiplication of logical identities, i.e. stabilisers. However it is this step where the cost of the simulation arises, both in time and space of simulation. Although this is not a physical update we will trace through what would happen in the computation to see how our update can be carried out to ensure consistency in maintaining commutation.

We first expand logical operators in terms of products of Pauli operators
\begin{equation} \label{EQN: Expansion of logical operators}
L_\alpha = \sum_i a_i M_i^\alpha,
\end{equation}
where $M_i^\alpha$ is some product of Pauli operators, this is always possible since the Pauli operators forms a complete operator basis. Then, starting in $R_1$ with the stabilisers (Equation \ref{EQN: Stabiliser group Step 1}) and logical operators (Equation \ref{EQN: Log Op Step 1}), we proceed with each round as follows, going from the first to the final round in sequence.
In round $R_x$ we update each Pauli term $M_i^\alpha$ in each $L_{\alpha}$ as follows:
\begin{eqnarray*}
\forall \mu \in R_x: \;\;\;\; &\text{If}& [M_i^\alpha,X_\mu]=0,\;\;\;\;\;\;\;M_i^\alpha \rightarrow M_i^\alpha \\
&\text{If}& \{M_i^\alpha,X_\mu\}=0, \;\;\;\;\;M_i^\alpha \rightarrow S_\mu M_i^\alpha
\end{eqnarray*}

After this step is complete, by the properties of gFlow it is easy to see that all the $L_\alpha$ will commute with all $X_\nu$, i.e. $[L_\alpha,X_\nu]=0$ $\forall \alpha, \forall \nu \in O_c$.

\textbf{Step 3:} The third step reflects the measurement of the computation, however with the unphysical condition that all outcomes are plus one. Although this does not really reflect measurement, it reflects the computation, since corrections are made so that this is always the final state. We first update the stabilisers and then use these to update the logical operators so that they are trivial (identity) everywhere except the outputs. The stabilisers are replaced with
\begin{equation} \label{EQN: Stabiliser group Step 3}
S=\left< \left\{X_i \right\}_{i \in O_C},\{G_i\}_i\in{O} \right>.
\end{equation}
One can picture this as measurements with fixed $+1$ outcome as follows. We first notice that $\{S_i,X_i\}=0$ and $[S_{i>j},X_j]=0$, as can easily be seen from the definition of gFlow. To update the stabiliser operators to arrive at Equation \ref{EQN: Stabiliser group Step 3} from Equation \ref{EQN: Stabiliser group Step 1}, again we start in $R_1$ and proceed with each round, going from the first to the final round, and in each round $R_x$, we replace all the stabilisers $S_{i\in R_x}$ with $X_i$ (corresponding to measuring $X_i$ and getting result $+1$). Because of the condition $\{S_i,X_i\}=0$ and $[S_{i>j},X_j]=0$, this reflects exactly measurement with the $+1$ outcomes, and finally we end up with the stabilisers (Equation \ref{EQN: Stabiliser group Step 3}).

The next part is to use these new stabilisers to update the logical operators one final time. Again we do so term by term in the decomposition into Pauli operators. If a term $M_i^\alpha$ has an $X_\mu$ for $\mu \notin O$, it is multiplied by $X_\mu$ (which is now a stabiliser, hence a logical identity). The remaining logical operators are trivial (i.e. identity) on everything except the outputs, and they encode the unitary evolution of the computation $L_\alpha \rightarrow U^\dagger L_\alpha U$. This completes the classical simulation.

\begin{figure}[h]
\centering
\resizebox{!}{4.3cm}{\includegraphics{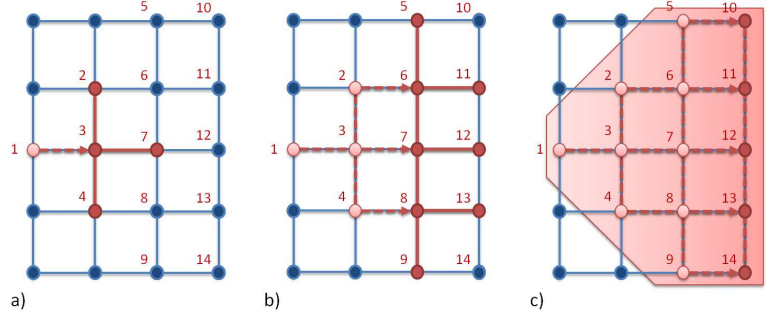}}
\caption{\label{FIG: UPdateCluster} Illustration of Step 2 the update procedure for the cluster state. The red vertices represent the qubits where the logical operators $L_{X_1}$ and $L_{Z_1}$ are non-trivial. a) is the point directly after $X_1$ has been considered. b) is the point directly after qubits in the first and second round have been measured. c) is the point after qubits in the third round have been considered. See text for details. The number of qubits touched in the update procedure is equal to the size of the forward cone $F_C$, which gives an upper bound to the size of the final logical operators, and hence the cost of directly simulating the computation (see observation \ref{OBS: gFlow simulation}). The $F_C$ also acts as a light cone for the information spread throughout the computation.}
\end{figure}

The efficiency of this procedure is dominated by the size of the logical operators (the number of terms occurring in the expansion). The stabilisers are updated efficiently (nothing in the initialisation or the update scales larger than $O(n)$ where $n$ is the size of the pattern). Similarly the initialisation of the logical operators is efficient, however, during each update step, each term in the expansion into Pauli operators must be checked and possibly updated. When an $S_\mu$ is added to the term in the second step, the size increases by $2^{|g(\mu)|}$, where $|g(\mu)|$ is the number of vertices in the correcting set. This is necessary for every Pauli $Z$ operator introduced by previous updates. Starting from $R_1$ these Pauli $Z$ operators are introduced on all the neighbours of the correcting sets - that is along the gFlow path and one graph edge further. Thus they follow along all possible influencing paths. Some terms may cancel out, so the total number of terms is less than equal to $2^{|F_C(\nu)|}$. From this we get the following observation.

\PRO \label{OBS: gFlow simulation} An MBQC over an open graph state with gFlow can be simulated classically in $O(n \exp(|F_{C_{max}}|))$ where $F_{C_{max}}$ is the maximum forward cone over all the inputs. More explicitly the logical operators ${L_\alpha}$ associated to vertex $\mu$ can be updated with $O(\exp(|F_C({\nu})|))$.
\ORP

As mentioned, the above simulation does not take into account correction (since it is unnecessary in terms of simulating the computation). One may wonder given the simulation above where would the corrections come in at all. The answer is in the last step - when measureing $X_i$, and getting result $r_i$, instead of replacing by $X_i$, we should replace by $-1^{r_i} X_i$. This would add signs throughout the logical operators which in general could not be undone by simply products of Pauli operators. With the exception being the case when each logical operator only has one $M^{\alpha}$  in expansion (Equation \ref{EQN: Expansion of logical operators}), i.e. is just a product of Pauli operators, which occurs when the angles $\theta_i=0,\pi$, i.e. measurements onto Pauli operators only. Then the minus signs can be all flipped coherently by multiplying by stabilisers. This is another way of seeing that if only Pauli measurements are made, all corrections can be made at the end. In such a case one can also see that the size of the logical operators becomes small - only one term each - so that this simulation itself is efficient. This simple observation will allow us to derive the equivalent of Gottesman-Knill Theorem directly in MBQC, as all the Clifford operates can be implemented in MBQC  using only Pauli operators. Having removed any dependency as described above will lead to an efficient classical simulation of any MBQC pattern implementing Clifford operators and Pauli measurements. This interplay between efficiency and the angles of measurement is something not taken into account in the above observation, and offers more potential for better bounds. We leave this to future work for now.

To see the updating which truly corresponds to a computation, i.e. including corrections, one can combine steps 2 and 3 to get rid of the $X_i$s round by round by applying the post measurement stabilisers $-1^{r_i} X_i$  and in addition perform the correction operation (given by the gFlow) to remove the $-1^{r_i}$. The effect is that one can simply remove the measured $X_i$s whilst tracing through the computation.

\bigskip

For clarity we go through the example for the first few rounds on the 2D cluster state.
For input of qubit $1$ before being attached to the graph it is described entirely by two logical operators $L_{X_1}=X_1$ and $L_{Z_1}=Z_1$. After Step 1 initialisation (joining to the open graph state and `rotating' each qubit according to the measurement basis), these become
\begin{align}
L_{X_1}&=e^{i \theta_1 Z_1} X_1 Z_3 \nonumber \\
L_{Z_1}&=Z_1. \nonumber
\end{align}
Here we have abbreviated the terms coming from the rotated basis into the exponent $e^{\theta_1 Z_1} = \cos \theta_1 \one + i \sin \theta_1 Z_1$, and for ease of notation we remove the tensor product symbol.

We next consider Step 2, starting with round $R_1$ and operator $X_1$ that anticommutes with $Z_1$s, hence for those terms in the $L_\alpha$ where this occurs we are required to multiply by $S_1= K_3^{\theta_3}=Z_1 \otimes Z_2 \otimes e^{i \theta_3 Z_3}X_3 \otimes Z_4 \otimes Z_7$. This is equivalent to putting it up into the exponent, so that the logical operators become
\begin{align}
L_{X_1}&=X_1 e^{i \theta_1 Z_2 e^{i \theta_3 Z_3}X_3  Z_4 Z_7}  Z_3 \nonumber \\
L_{Z_1}&= Z_2  e^{i \theta_3 Z_3}X_3  Z_4  Z_7. \nonumber
\end{align}

In Step 3 the $X_1$s are removed (since after measurement and correction the $X_1$ are a logical identity), and the logical operators are thus non-trivial on qubits $2,3,4,7$ after $R_1$, as illustrated in Figure \ref{FIG: UPdateCluster}a). In the second round $R_2$,  $X_\nu$ on qubits $2$, $3$ and $4$ are considered. We update the logical operators by considering these one by one, starting with $X_2$ (any order in the same round gives the same final result). This anti commutes with $Z_2$ - which comes from the application of $S_1 = K_3^{\theta_3}$ in the previous round. Indeed this is how the updates are effected along all influencing paths. When the $Z_2$ occurs we are forced to multiply the term by $S_2=K_6^{\theta_6} = Z_2 \otimes Z_5 \otimes e^{\theta_6 Z_6} X_6 \otimes Z_7 \otimes Z_{11}$. This takes the logical operators to
\begin{align}
L_{X_1}&=e^{i \theta_1 e^{i \theta_3 Z_3}X_3 Z_4 Z_5  e^{\theta_6 Z_6} X_6  Z_{11} } Z_3 \nonumber \\
L_{Z_1}&= e^{i \theta_3 Z_3}X_3  Z_4  Z_5  e^{\theta_6 Z_6} X_6  Z_{11} . \nonumber
\end{align}
Note that here, two $Z_7$ operators have cancelled out - they came from two occasions where qubit $7$ was a neighbour of one of the correcting sets. In this way, it is possible that some qubits in the set of influencing paths cancel out - this happens if the number of influencing paths they sit in as gFlow paths is even, and the number arriving from non-gFlow paths is also even (at this point in our calculation the number of times it is on a gFlow path is zero, and it is in 2 influencing paths not as a gFlow).

After qubits $X_3$ and $X_4$ are also considered, we have to do the same trick to get rid of the $Z_3$ and $Z_4$s, by multiplying the terms where they occur by $S_3=K_7^{\theta_7}$ and $S_4=K_8^{\theta_8}$ respectively. Finally we end up with logical operators

\begin{align}
L_{X_1}&=e^{i \theta_1 e^{i \theta_3 Z_6 e^{i \theta_7 Z_7}X_7 Z_8 Z_{12}} X_3   Z_5  e^{\theta_6 Z_6} X_6 Z_7 e^{i \theta_8 Z_8}X_8 Z_9  Z_{11}Z_{13}} Z_6 e^{i \theta_7 Z_7}X_7 Z_8 Z_{12} \nonumber \\
L_{Z_1}&= X_3 e^{i \theta_3 Z_6 e^{i \theta_7 Z_7}X_7 Z_8 Z_{12}  } Z_5  e^{\theta_6 Z_6} X_6  Z_7 e^{i \theta_8 Z_8} X_8 Z_{11} Z_{12}  . \nonumber
\end{align}

Again, in Step 3 we get rid of the $X_2$, $X_3$, $X_4$s, hence after the measurements in round $R_2$ the logical operators are non-trivial only on qubits $6,7,8,9,11,12,13$, as indicated in Figure \ref{FIG: UPdateCluster}b). It is then clear how after the third round of measurements we will be left 	with logical operators that sit on the highlighted qubits $10$, $11$, $12$, $13$ and $14$, as indicated in Figure \ref{FIG: UPdateCluster}c).

For any graph and any measurement pattern with gFlow, each time a Pauli $Z$ operator is added, unless it is in the output set, we will have to multiply that term by a stabiliser - which will add a splitting of two. During the update procedure, Pauli $Z$ operators are added along every influencing path. Sometimes these will cancel out, depending on the graph, but sometimes not, so that this gives an upper bound to the complexity for the direct update procedure which is the content of Observation \ref{OBS: gFlow simulation}.

We thus see an initial way to go from a gFlow to a classical simulation. However, for certain examples this bound can be very bad. We have already mentioned that this is the case  where all of the measurement angles are zero or $\pi/2$ - i.e. measuring the Pauli operators - there is no splitting of the logical operators, and only one term is needed for each logical operator, hence this simulation becomes efficient, which is not captured by Observation \ref{OBS: gFlow simulation} (where we effectively assume the worst case for the angles). Another example is a computation across along a 1D graph state, with one input, say on the left, and an output on the right (see Fig. 3c). There the gFlow simply follows the line, thus the influencing volume is big, however, this is always a simple one qubit computation, and indeed all computations on a 1D cluster state are easy to simulate classically \cite{Nielsen05}. In Section \ref{SEC: Ent Simulation Bound} we will see how connections to entanglement allow us to make tighter bounds on classical simulatability which will work well for this example and many others. Before we do that however, in the next section we will discuss how the update above can be interpreted as information flow, in tern giving the interpretation of the forward cone $F_C$ as a light cone for the information.

\section{Flow of Information and $F_C$ as Information Light Cones} \label{SEC: InfoFlow}

The gFlow gives a causal structure on top of a graph state induced by the correction procedure, called the  forward cone  $F_C$ (Definition \ref{Forward Cone}). In this section we will also look at how the same cone can be understood as a forward cone of information,  and moreover a light cone (so that information cannot travel beyond this cone).

The forward cone can be viewed as an information forward cone directly from the simulation procedure described in the previous section, and the interpretation of the logical Heisenberg representation as showing us where information sits (see for example \cite{DH99}). Consider a density matrix of some input $i$
\begin{align} \label{EQN: density Matrix}
\rho_i = \frac{1}{2}\left(\one + \eta_x X_i + \eta_y Y_i + \eta_z Z_i\right)
\end{align}
The state is totally described by the coefficients $\eta_i$. The logical Heisenberg representation ensures that at any time the evolved state, denoted as $\tilde{\rho}$, which now can be sitting over many systems, is described as
\begin{align}\label{EQN: density Matrix Update}
\tilde{\rho} = \frac{1}{2}\left(\one + \eta_x \tilde{L_{X_i}} + \eta_y \tilde{L_{Y_i}} + \eta_z \tilde{L_{Z_i}}\right),
\end{align}
where the $\tilde{L_\alpha}$ are the updated logical operators of $\alpha$ corresponding to the evolution.

The information is then preserved, but `spread' over to different qubits in the following sense.
To recover the information encoded on the original system $i$ (i.e. recover the $\eta_i$), we should measure the logical operators $\tilde{L_{X_i}}$, $\tilde{L_{Y_i}}$ and $\tilde{L_{Z_i}}$. Thus the information can be said to have spread over the range of the logical operators. From the simulation in the previous section, it is clear that the logical operators, and hence the information of input qubit $i$ spreads out over the causal forward cone $F_C$ defined by the gFlow (see Figure \ref{FIG: UPdateCluster}).

One may then ask if this is all that is allowed, or could we understand the information as having spread further than the influencing cone (after all, this is not the only way one may simulate a computation)? The answer (at least for patterns where we wish uniform determinism, i.e. that all measurements on a Pauli equator are allowed) is no, in that the spread must be balanced by consistency amongst all measurements, which is the function of the gFlow, which defines the cone $F_C$.

To see how this works, let us first return back to Step 2 in the simulation above, which is where this spread of information occurs in the simulation. The trick is simply multiplying the logical operators by a logical identity (i.e. the stabilisers). This part however is clearly not restricted to the cone. One could easily expand a logical operator to cover practically all qubits in this way. The reason we do not allow this is because we want to do measurements, and we want to do them over all qubits not outputs so that all logical operators are preserved (this is what we mean by consistency). Say one did this for operator $L_{X_i}$, so that its extent was over many qubits. Taking its expansion into products of Pauli operators (as in Equation \ref{EQN: Expansion of logical operators}), one would have a sum of many terms, including Pauli $Z$ and $X$ operators on any given qubit in its range.
When measuring qubits, to ensure the survival of the information, we asked that the logical operators be taken to a form which commutes with the measurement - this was the role of Step 2 in the above.
If one did not have this, information may be lost.
This can only be the case if in each term $M_j^{X_i}$ of the expansion of $L_{X_i}$, the part of $M_j^{X_i}$ on vertex $\mu$ is the same (say $\sigma_\mu$) or identity for all terms.

 One could have, for example, that this is indeed the case, i.e. for a particular $L_{X_i}$, extended so that it touched many qubits, that over each such qubit $\mu$ all the terms in the expansion of $L_{X_i}$ were either the same Pauli $\sigma_\mu$  or identity. In such a case, one could happily measure those qubits in the Pauli basis $\sigma_\mu$, the information would be preserved, and the logical operators could be calculated (if we wanted to consider the information over the outputs we would then follow Step 3 to leave them as identity everywhere else, though one would have potentially different evolutions for different branches). In this way its final spread may indeed be beyond the light cone given by gFlow. The problem with this would be that we want to transfer the logical operators not just of one input $i$, but of \emph{all} the inputs. It is shown in \cite{Browne06} that to achieve this, in such a way that every measured vertex one can choose amongst a set of measurements across one of the planes, the only way to do it is via a gFlow. Hence, for an input $i$, not only is the forward cone $F_C(i)$ also an information cone, but to transfer all the information at the same time, it is a light cone for the information contained $i$ - that is, the information can not spread beyond it, and the computation be consistent for all inputs.

From the perspective of information flow, observation 2 says, unsurprisingly, that the more information is spread through a computation, the more costly it is to simulate. However, again we should be careful to note that the true cost of simulation depends on the angles of the measurements, which is not captured by the size of $F_C$, hence not by observation 2. As we saw, for angles $0,\pi$, the simulation is simple, however the spread of information is still large.

\section{Bounds on Entanglement from gFlow} \label{SEC: Ent Simulation Bound}

In this section we will show a connection between gFlow on the one hand, and entanglement conditions for both the universality of a resource state and the classical simulatability of a computation on the given resource, on the other. More precisely we will show that the Flow and gFlow can be used to bound the entanglement of the graph state, in terms of the entanglement width  \cite{VandenNest06} and what we will call the \emph{structural entanglement} (though not explicitly defined, it can be understood from \cite{Yoran06}, see also \cite{Vidal03b}). Conditions of universal family of resource states, and for classical simulatability are known for both these measures, which can be translated to conditions about the gFlow through our bounds \cite{VandenNest06,VandenNest06b,VandenNest07}. Several known results can then be derived for both the measurement based model and circuit based model (through the known maps between the two models \cite{Danos06,Beaudrap09}). For example we reproduce the result by Josza stating that a circuit which any wire are touched by at most logarithmic (in the size of the input) many number of two qubit gates can be classically simulated efficiently \cite{Jozsa06}.

Let us first define the entanglement measures we are interested in. The entanglement width \cite{VandenNest06} of a pure state $|\psi\rangle$ is defined as
\begin{eqnarray}
\chi_{wd}(|\psi\rangle):= \min_T \max_e \chi^{bi}_{T,e}(|\psi\rangle),
\end{eqnarray}
where $\chi^{bi}_{T,e}(|\psi\rangle)$ is the the $\log$-Schmidt rank across the bipartite cut defined by $T$ and $e$ where $T$ is a sub-cubic graph with $n$ leaves and $e$ is an edge of $T$. Each leaf corresponds to a qubit of the state $|\psi\rangle$. The bipartite cut is defined by removing edge $e$ to give two separate trees. The leaves of one tree correspond to one side of the cut, and the other tree the other side of the cut. It was shown that if the entanglement of a family of resource state does not scale polynomially with the size of its input space then, that family cannot be a universal \cite{VandenNest06,VandenNest07} (in the case of QQ computations, note that this is not the same as asking for universality in the CC case). It was also shown that any MBQC can be simulated in $O(n poly(2^{\chi_{wd}}))$ \cite{VandenNest06b}.

Motivated by the proofs in \cite{Yoran06}, we define the \emph{structural entanglement} as
\begin{eqnarray}
E_{struc}(|\psi\rangle) := \min_{\substack{Order\\ 1 ,\ldots, n}} \; \;\max_{\substack{cut\;k \\A=1 ,\ldots, k \\B=k+1 ,\ldots, n}} \chi_{AB}(|\psi\rangle),
\end{eqnarray}
where the minimum is taken over all orderings (labelings) of the qubits $1 ,\ldots, n$, and the max is taken over a cut defined for a given ordering by taking all qubits $1 ,\ldots, k$ on partition $A$ and the rest on partition $B$ and $\chi_{AB}(|\psi\rangle)$ being the $\log$-Schmidt rank over cut $AB$. Although not explicitly stated in terms of this measure, in \cite{Yoran06} it is shown that any MBQC pattern can be simulated classically in $O(n^2 poly(2^{E_{struc}}))$.

\begin{figure}[h]
\centering
\resizebox{!}{4.3cm}{\includegraphics{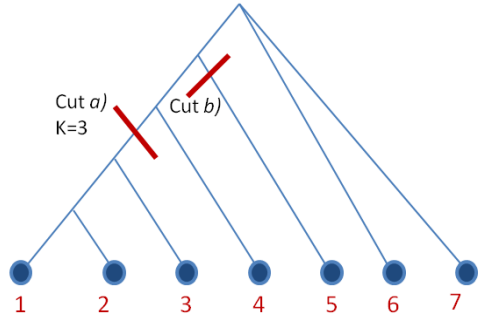}}
\caption{\label{FIG: Tree} This tree defines a set of cuts showing that $E_{struc}\geq \chi_{wd}$.
Any cut either sits in the same set of cuts as that optimised for $E_{struc}$ - effectively choosing a $k$ such that all qubits of number less than or equal to $k$ are in partition $A$ and all higher qubits are in partition $B$ (cut a)), or else it singles out one qubit (cut b)), which can never be the unique maximum.}
\end{figure}

It is easy to see that the tree in Figure \ref{FIG: Tree} defines a set of cuts such that any cut either splits the graph in two with all leaves below or equal to a value $k$ on the left, and above $k$ on the right (as per the optimisation for $E_{struc}$), or else it just identifies one leaf. This clearly implies that
\begin{equation}
E_{struc}\geq \chi_{wd}.
\end{equation}

We will now see how the Flow and gFlow can be used to bound $E_{struc}$, and in turn $\chi_{wd}$. We start by considering Flow, which is simpler to picture, but the ideas easily extend to gFlow. The idea is that they both can be used to define a natural order, which gives a simple bound to $E_{struc}$ which comes from induced disjoint input-output paths. Indeed, if an open graph state has Flow, following the image of the Flow function, $f$, (Definition \ref{g-flow}) from each of the inputs  leads to disjoint lines to the outputs, which cover all the non outputs {\cite{Beaudrap06,Beaudrap09}} (called \emph{Flow wires}).

\begin{figure}[h]
\centering
{\resizebox{!}{6cm}{\includegraphics{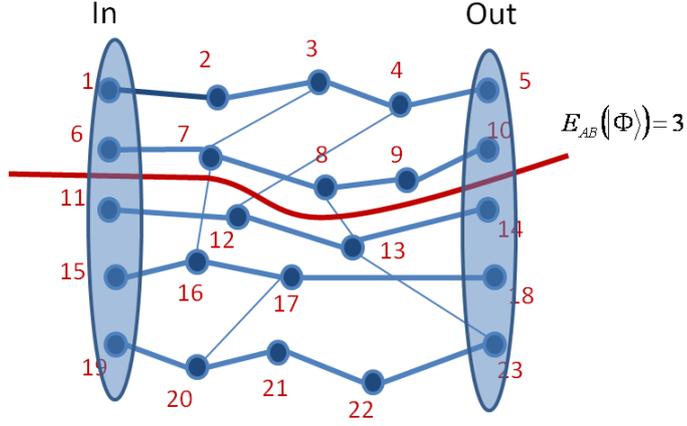}}} \caption{\label{FIG: Estruc} Flow defines a natural ordering from top left to right across each Flow wire from top to bottom as shown. This is used to define cuts by a number $k$ where partition $A$ consists of all qubits below qubit $k$ in the ordering and partition $B$ consists of all qubits above $k$ in the ordering. The entanglement across any cut is upper bounded by the number of edges cut (in this case $k=10$ and the entanglement is exact).}
\end{figure}

\begin{figure}[h]
\centering
{\resizebox{!}{4.3cm}{\includegraphics{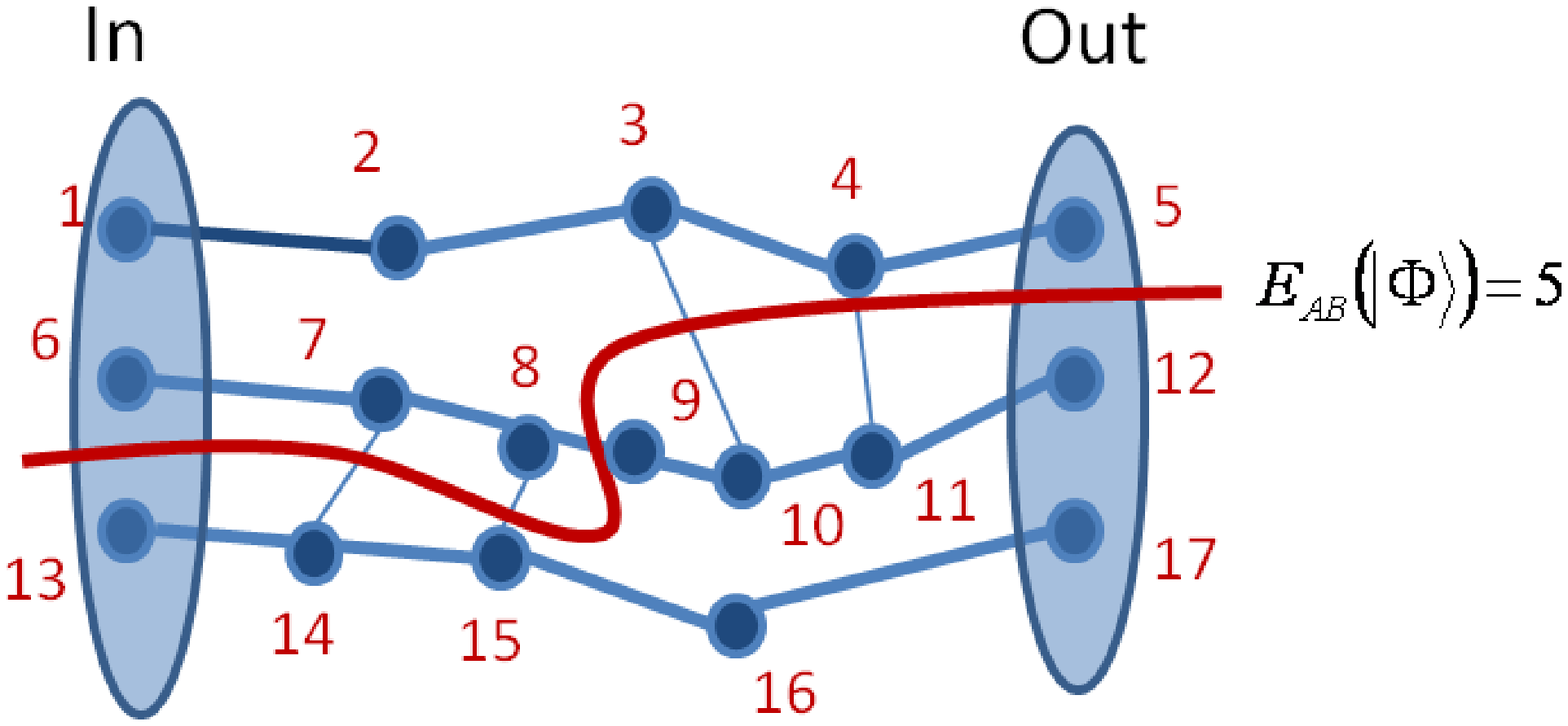}}} \caption{\label{FIG: Estruc2} In the worst case the number of edges cut by a line equals $1+2C_F$. This implies a bound on the structural entanglement (see text).}
\end{figure}

We consider first the case where $|I|=|O|$. The numbering goes as follows. We start with an arbitrary input going along the image of the Flow function of that vertex till we reach an output qubit. Next we choose another not selected input qubit and carry on in this fashion, till we cover all the inputs, see Figure \ref{FIG: Estruc} and Figure \ref{FIG: Estruc2}. Note that based on the definition of Flow, no input qubit could belong to the image of the Flow function of another input qubits hence on each such Flow wire there will sit only one input qubit and hence we have $|I|$ such wires. To calculate the entanglement we note the fact that the entanglement across a cut $C$ for a graph state can be bounded by the number of edges crossing the cut $E_{A,B}\leq C$. This is clear since in preparation of the state each edge corresponds to a control-$Z$ operator, and $C$ such operators can create at most $C$ e-bits.
\DE \label{C_E} For an open graph with flow, we denote $C_F$ the maximum number of edges crossing between Flow wires.
\ED

We easily see that a cut between two Flow wires gives entanglement at most $C_F$ (see Figure \ref{FIG: Estruc}). This can be at most doubled (plus one) by choosing a lower number to cut at (thus potentially increasing the number of edges cut) (see Figure \ref{FIG: Estruc2}). We thus have that $E_{struc}\leq 1+ 2C_F$.

To extend this to the case where $|O|>|I|$ we must consider the worst case, for which each extra qubit adds one unit of entanglement. In this general setting we now call $C_F$ the maximum number of edges crossing between Flow wires, when the output qubits not in a Flow wire are ignored along with their edges. We also call $\Delta:=|O|-|I|$. We then have the following observation.

\PRO \label{OBS: Ent Flow simulation}
A graph state with Flow has structural entanglement
\begin{eqnarray}
E_{struc}\leq 1+ 2C_F + \Delta.
\end{eqnarray}
Thus any computation can be simulated in at least  $O(n^2 poly(2^{2C_F+\Delta}))$.
\ORP
We note that any computation which can be done with a number of outputs greater than $|I|$ can be done with $|I|=|O|$ without changing the Flow or $C_F$ by simply removing the extra $\Delta$ output qubits from the graph resource. Thus $\Delta=0$ for most interesting cases. This is clear since the existence of Flow is robust against losing the extra outputs, and this guarantees the computation.

This result can be extended to open graphs with gFlow by using gFlow to find disjoint input-output lines as follows. As we saw earlier, it is clear that if an open graph state has gFlow, then it is necessarily $D$-happy (from Observation \ref{t-dhappy} and the fact that gFlow implies unitary computation). This in turn means that  any cut which separates the input and the output goes through at least $C\geq |I|$ edges. Taking a result from graph theory, Menger's theorem \cite{Diestel} says that this implies there are at least $n$ parallel paths going from inputs to outputs. Furthermore it can be shown that there are parallel paths which sit along gFlow paths and can be found systematically also \cite{MHM13}. This can be used to give a natural order to the graph as for Flow, but with the possibility that non-output qubits do not sit in the disjoint paths, and so should be added to the $\Delta$ term. Again the size of $\Delta$ may be reduced or removed by considering equivalent smaller graphs, but this is less well understood for gFlow.

This result covers examples not covered by the direct simulation from Section \ref{SEC: Direct simulation}, for example the $1D$ cluster state. The statement of Observation \ref{OBS: Ent Flow simulation} is a very similar sounding statement to Jozsa's \cite{Jozsa06} which states that a quantum computation on a circuit can be simulated in $O(n poly(2^D))$ where $D$ is the maximum number of gates that touch or cross a circuit wire.

\section{Conclusions}

We have seen that gFlow can be used for two complementary approaches for giving bounds on the efficiency of classical simulation for MBQC. In Observation \ref{OBS: gFlow simulation} we saw that classical simulation is possible with resources scaling as exponential in the size of the largest causal future cone defined by the gFlow. On the other hand in Observation \ref{OBS: Ent Flow simulation} gFlow can be used to bound the entanglement, and hence give bounds on resources for classical simulation in terms of the number of edges crossing gFlow wires (parallel wires from input to output induced by gFlow). Simple and straightforward, but illuminating conditions for entanglement of general resource states are described in Observation \ref{t-dhappy}. Furthermore the causal future cone induced by gFlow is seen to be at the same time a light cone for information spreading.

The results on classical simulation combine two of the main approaches for bounding the cost of classical simulation for quantum computation - explicit tracking of the computation using an efficient form (used for example in the Gottesman Knill \cite{Gott97} theorem and related results (e.g. \cite{CJL07})) and bounds coming from entanglement (used for simulating computation \cite{Vidal03b,Yoran06,VandenNest06,VandenNest06b,VandenNest07} as well as many body physics (e.g. \cite{Vidal08})). This offers the perspective of bridging these two approaches via gFlow. A natural question is the interplay between the angle of measurement and efficiency of simulation via the gFlow update procedure presented here. Setting all angles to zero or $\pi$ makes the simulation efficient (as per the Gottesman Knill theorem), however for general angles it is not efficient(indeed Observation \ref{OBS: gFlow simulation} represents this worst case situation). The in between ground, combined with bounds by entanglement may present new classes of computation admitting efficient simulation for example. Furthermore, we may gain more insight into how efficiency of classical simulation is related to other features of computation illuminated by the study of gFlow.

It is also interesting in itself that from Flow and gFlow one can derive bounds on the entanglement of a graph state. Since there exist efficient algorithms to calculate the Flow and gFlow of graphs \cite{Beaudrap06,MhallaP08}, and given Flow and gFlow one can easily bound the entanglement, one may use this to bound the entanglement of general graph states. This is both important for recognising good resources (since the existence of Flow does not talk about universality, whereas the entanglement gives bounds on this also \cite{VandenNest06,VandenNest07}), and more generally as entanglement represents important resource in other areas of quantum information.

The two approaches to classical simulateability can also be understood as arising from two notions of `spreading' of information. We have seen in Section \ref{SEC: InfoFlow} that the forward cone given by gFlow bounding the cost of classical simulation can also be interpreted as a spread of information - so that the more spread the information is, the more costly the simulation. The bounds arising from entanglement (\cite{Vidal03b}, \cite{Yoran06} e.t.c) which lead to Observation \ref{OBS: Ent Flow simulation}, can also be understood as assigning a cost to the spread of information as follows. The entanglement measure key to these results is a bipartite measure, the Schmidt measure of entanglement, which counts the minimum number of product states (with respect to a particular cut) needed to describe the state. This may be interpreted as saying how `spread' across product bases the state is. Indeed it is exactly the rank of the reduced density matrix of one cut, so in a sense says the size of the space in which it must be understood to sit (in this sense the `spread' is over the state space rather than precisely the parties). The trick of \cite{Vidal03b} and subsequent work is to find an efficient form to describe the state and its updating through a computation based on this minimum decomposition. Again, the smaller the `spread' in this sense, the smaller the cost of this simulation. As we have also seen in Sections \ref{SEC: Direct simulation} and \ref{SEC: InfoFlow}, a big `spread' of information is however not enough to imply that a computation is difficult to simulate - MBQC with only Pauli measurements is efficient to simulate, but the spread of information is large (in both senses - the future cone is large, and the entanglement is large). To capture the difficulty in simulation, one must also include something about how this `spread' of information is used. In the case of MBQC studied here, universality (and presumably the difficulty in simulation) is given by the use of arbitrary angles for the measurements, using the spread of information in the most universal way. This balance between spread of information through entanglement and how it can be used also plays a key role in analogies between MBQC and thermodynamics and in particular phase transitions \cite{Anders07,Markham10}. It is an exciting prospect that these pictures may be unified from the perspective of gFlow or similar notions.

\noindent ACKNOWLEDGEMENTS\\
The authors would like to thank Bobby Antonio, Simon Perdrix and Einar Puis for useful discussions and feedback. We are particularly grateful to Vincent Danos and Prakash Panangaden for many discussions on the topics of this paper which gave rise to many of the ideas mentioned here directly and indirectly. DM is funded by the FREQUENCY (ANR-09-BLAN-0410), HIPERCOM (2011-CHRI-006) projects, and by the Ville de Paris Emergences program, project CiQWii. EK is funded by UK Engineering and Physical Sciences Research Council grant EP/E059600/1.

\bibliographystyle{unsrt}
\bibliography{MBQC}

\end{document}